\documentclass{aa}
\usepackage{epsfig}
\def\ltsima{$\; \buildrel < \over \sim \;$}
\def\simlt{\lower.5ex\hbox{\ltsima}}
\def\gtsima{$\; \buildrel > \over \sim \;$}
\def\simgt{\lower.5ex\hbox{\gtsima}}

\begin{document}
\thesaurus{}
   \thesaurus{5(11.01.2;  
		11.02.2 Mkn~421; 
                11.10.1;  
                11.14.1;  
                13.25.2)}  

\title{The quiescent state broadband X-ray spectrum and variability of Mkn~421}

\author{M. Guainazzi\inst{1}, G. Vacanti\inst{1}, A. Malizia\inst{2,3}, K.S. O'Flaherty\inst{1}, E. Palazzi\inst{4}, A.N. Parmar\inst{1}}

\institute{
{Astrophysics Division, Space Science Department of ESA, ESTEC, Postbus 299,
NL-2200 AG Noordwijk, The Netherlands}
\and
{BeppoSAX Science Data Center, Via Corcolle 19, I-00131 Roma, Italy}
\and
{Department of Physics and Astronomy, Southampton University, SO17 1BJ, United Kingdom}
\and
{Istituto per le Tecnologie e Studio delle Radiazioni Extraterrestri, CNR, Via Gobetti 101, I-40129 Bologna, Italy}
}
   
\offprints{M.Guainazzi (mguainaz@astro.estec.esa.nl)}

\date{Received 6 July 1998; accepted }

\maketitle

\markboth{M.Guainazzi et al.}{The quiescent state broadband X-ray spectrum and variability of Mkn~421}

\begin{abstract}

The BL Lac object Mkn~421 was observed three times
by the X-ray observatory BeppoSAX
in consecutive days during 1997 April and May.
The source was in a quiescent state,
with an average
2--10~keV flux of $9.0 \times 10^{-11}$~erg~cm$^{-2}$~s$^{-1}$.
Flux variation by a factor of $\simeq$2 on timescales as
short as a few $10^4$~s were more pronounced in the hard (i.e. above
$\simeq$3~keV) than in the soft X-rays.
The broadband (0.1--40~keV) spectrum is concave and can
be most easily explained with a power-law model which
steepens gradually with energy. In this
framework, neither photoabsorption edges nor resonant absorption lines
are required, strengthening the case against the ubiquity of
such features in BL Lac objects, which had been previously suggested by
{\it Einstein} observations.
The broadband spectrum hardens with hard X-ray flux, mostly due
to a flattening above $\simeq$4~keV. This suggests
that the relativistic highest energy electron distribution
properties drive the X-ray spectral
dynamics: either a stratification of the distribution in the jet
with energy or inhomogeneities in the electron injection mechanism
could be consistent with the observed variability pattern.

\end{abstract}
  
\keywords   {Galaxies: active --
		BL Lacertae objects: individual: Mkn~421 --
                Galaxies: jet --
                Galaxies: nuclei --
                X-rays: galaxies}

\section{Introduction}

Mkn~421 was the first BL Lac object to be detected in X-rays
(Ricketts, Cooke \& Pounds 1976). Subsequent observations
have detected the source up
to 100~keV ({\it e.g.}: Ubertini et al. 1984).
The X-ray emission of Mkn~421 is highly variable, with distinct
behavioral differences between soft and hard X-rays.
In the 0.5-10~keV band the source is variable on time
scales ranging from several hours to days, occasionally exhibiting
large X-ray outbursts characterized by a marked hardening of the
spectrum. EXOSAT observations showed that during the quiescent state
the source hardens as it brightens
(George, Warwick \& McHardy 1988). Giommi et al. (1990) showed that flux variations
in the hard X-ray band (0.7--8~keV) were more pronounced than those in
the soft band (0.06--0.3~keV). 

Observations with EXOSAT (Giommi et al. 1990), ROSAT (Fink et al. 1991)
and ASCA (Kubo et al. 1998) reveal that variability of Mkn~421
in the soft X-ray band can be
represented by small-amplitude variations about a quiescent level that
remains temporally relatively constant. The timescale of these
variations is of the order of days to weeks (George et al. 1988).

Mkn~421 has also been detected in ${\rm \gamma}$-rays by Compton
GRO/EGRET (Lin et al. 1996), and it was the first extragalactic
source to be detected at TeV energies (Punch et al. 1992).
Multiwavelength observations have shown that variability in the X-ray
band is accompanied by significant variability in the TeV region,
with little or no change in other wave bands (Macomb et al. 1995).

In this paper we report on three observations of Mkn~421 in
1997 April and May with the Narrow Field Instruments (NFI) on board the
BeppoSAX X-ray observatory (see Table~\ref{tab1} for a log
\begin{table*}[htbp]
\caption{Mkn~421 observations log. Note that the LECS is only operated
during spacecraft nighttime and collimated detectors spend about 
$50\%$ of the time observing the background}
\begin{center}
\begin{tabular}{lcccccc}
& \multicolumn{2}{c}{1997 April 29 (MJD 50567)} & \multicolumn{2}{c}{1997 April 30} & \multicolumn{2}{c}{1997 May 1} \\ \hline \hline
Instrument&Count rate &  Exp. time &Count rate &  Exp. time&Count rate &  Exp. time \\
           & (Hz)        & (ks)          & (Hz) & (ks) & (Hz) & (ks) \\ \hline
LECS & 2.119 $\pm$ 0.0137 & 11.6 & 2.097 $\pm$ 0.0138 & 11.4 & 2.529
 $\pm$ 0.0153 & 11.2\\
MECS1 & 0.544 $\pm$ 0.005 & 21.9 & 0.535 $\pm$ 0.005 & 24.0 & 0.633
 $\pm$ 0.005 & 23.8 \\
MECS2 & 0.797 $\pm$ 0.006 & 21.9 & 0.783 $\pm$ 0.006 & 24.0 & 0.915
 $\pm$ 0.006 & 23.8 \\
MECS3 & 0.827 $\pm$ 0.006 & 21.9 & 0.821 $\pm$ 0.006 & 24.0 & 0.970
 $\pm$ 0.006 & 23.8 \\
PDS &0.09 $\pm$ 0.03 &11.0 &0.19 $\pm$ 0.03 &10.8 &0.16 $\pm$ 0.03
&9.1 \\ \hline \hline
    \end{tabular}
    \label{tab1}
  \end{center}
\end{table*}
of the observations).

\section{Observations and data reduction}

The Italian-Dutch BeppoSAX X-ray observatory (Boella, Perola \& Scarsi 1997a)
carries four co-aligned
NFI, which cover more than three decades of energy from 0.1 to
300~keV. These are:

\begin{itemize}
  
\item the Low Energy Concentrator Spectrometer (0.1--10~keV, LECS,
  Parmar et al.  1997);
  
\item three identical Medium Energy Concentrator Spectrometers
  (1.3--10~keV, MECS, Boella et al. 1997b);

\item a High Pressure Gas Scintillator Proportional Counter
  (4--120~keV, HPGSPC, Manzo et al. 1997).
  
\item a collimated Phoswitch Detector System (13--300~keV, PDS,
  Frontera et al.  1997);

\end{itemize}

In this paper results from the LECS, MECS and PDS instruments are
presented, because the HPGSPC data could not provide any significant
spectral constraint. The HPGSPC is actually tuned for spectroscopy of bright
sources with good energy resolution, while the PDS possesses an
unprecedented sensitivity in its energy band.
Data were processed using
the SAXDAS package (version 1.3.0).
Publicly available response matrices (September 1997 release) were used.

For the imaging detectors, source spectra were extracted from regions
of 8$^{\prime}$ and 6$^{\prime}$
radius, centered on the source positions in the
LECS and MECS, respectively. Background
spectra were extracted from blank sky fields.
The background contributes less
than 0.1\% and 1\% to the LECS and MECS full bandpass count rates of Mkn~421,
respectively, and less than 10\% in each energy channel
even at the highest MECS bandpass energies.
Background subtraction is therefore not critical for such a bright
source.

PDS spectra were
accumulated after excluding interval of source eclipse
and each 5 minutes after a South Atlantic Geomagnetic Anomaly
Passage, to avoid a gain instability due to recovery to the
nominal crystal voltages after switch-on.
PDS events were screened with a temperature-dependent
Rise Time (RT) threshold, which allows a reduction by a factor
of up to 50\% of the instrumental background.
Mkn~421 is detected at 8$\sigma$ level in the 13--30~keV
band.
All spectra were rebinned in order to have at least 20
counts per channel, to ensure the applicability of $\chi^2$ statistics.
In the following, energies are quoted in the source frame and uncertainties
at 90\% level of confidence for each parameter
(i.e.: $\Delta \chi^2 = 2.71$, Lampton, Margon \& Bowyer 1976),
unless otherwise specified.

\section{Results}

\subsection{Timing analysis}

\begin{figure}
\begin{center}
\epsfig{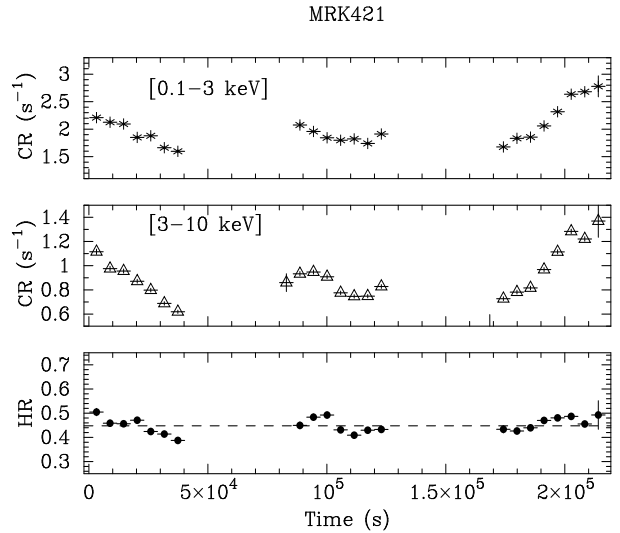}
\end{center}
\caption{Light curves in the 0.1--3.0~keV (LECS, upper panel),
3.0--10.5~keV (MECS, central panel) and their hardness ratio (HR).
Binning time is 5700~s, approximately one BeppoSAX orbit. The
dashed line marks the average HR}
\label{fig1}
\end{figure}

\begin{figure}
\begin{center}
\epsfig{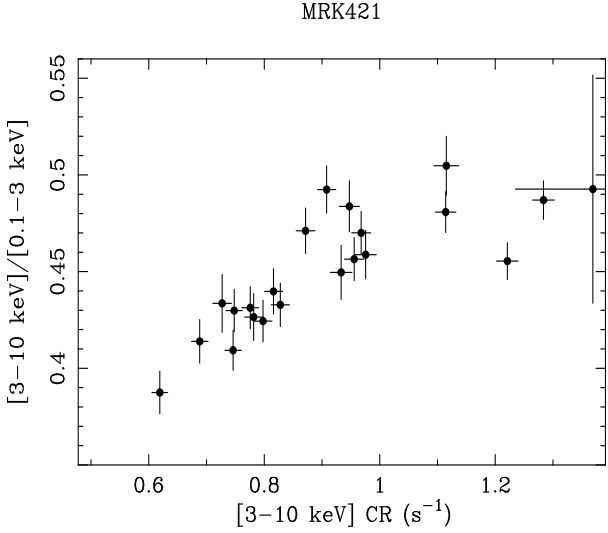}
\end{center}
\caption{Hardness ratio constructed from the 3--10~keV and 0.1--3~keV 
counts versus 3--10~keV
count rate. Each data point corresponds to an integration time
of 5700~s}
\label{fig2}
\end{figure}

In Fig.~\ref{fig1} the satellite orbit--averaged
(i.e.: binning time 5700~s)
light curves in the 0.1--3~keV (LECS) and 3--10~keV (MECS) energy ranges are
shown (the energy boundaries have been chosen to sample
different spectral components, see Sect.~3.2).
Peak-to-peak variability by factors of 80\% and 130\% is
evident in the 0.1--3~keV and 3--10~keV bands, respectively. The
3--10~keV/0.1--3~keV hardness ratio (HR) exhibits
a much smaller ($\simeq 20\%$) dynamical range.
Fig.~\ref{fig2} shows HR plotted against intensity (CR)
and shows that the HR tends to increase with increasing intensity
up to ${\rm CR_{break}}$ and then ``saturates'' to
a constant level. If the data points in Fig.~\ref{fig2} are fit with
a broken linear relation, ${\rm CR_{break} \simeq 0.82}$~s$^{-1}$.

\begin{figure}
\begin{center}
\epsfig{figure=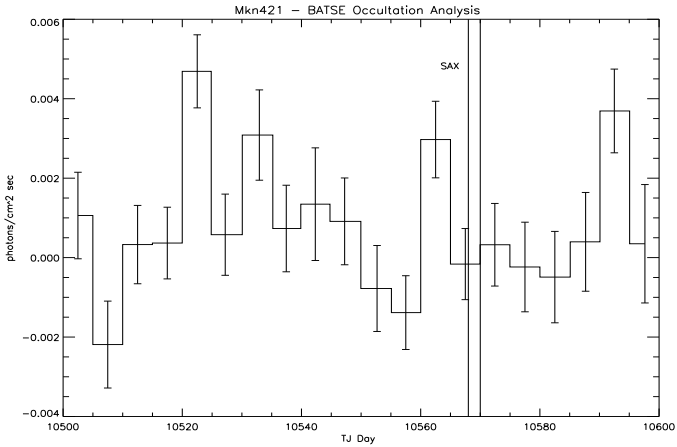,height=9.0cm,width=6.0cm,angle=-270}
\end{center}
\caption{BATSE light curve of Mkn~421 for the 100-day period covering
the three days of BeppoSAX observation,
using a bin of 5 days of the daily data. The BATSE data were
collected by the Large Area Detectors in Earth occultation mode (Harmon et al. 1992). Fluxes correspond to the 20--100~keV range
and have been calculated by folding a simple power-law of
photon index 1.7}
\label{fig3}
\end{figure}

This pattern of variability is typical of the quiescent X-ray state of
Mkn~421 (Giommi et al. 1990; Sambruna et al. 1994).
Indeed, simultaneous optical and TeV ${\rm \gamma}$-ray observations of
the source confirm the lack of any significant activity at the time
(McEnery \& Weekes, private communication).
The state of quiescence is further substantiated by the BATSE light
curve covering the the three days of the
BeppoSAX observations (see Fig.~\ref{fig3}).
The BATSE daily monitoring sensitivity is $\sim$100~mCrab, implying
that sensitivities of the order of few mCrab could be achieved by
integrating the data over a period of few years.
Over the 100 days covering the BeppoSAX observation, the 2$\sigma$
upper limit on the 20--100~keV flux was
${\rm 11 \times 10^{-11}}$~erg~s$^{-1}$~cm$^{-2}$.

\subsection{Spectral analysis}

Spectra from the three observations were combined, in
order to maximize the signal statistics. However,
given the spectral variations revealed by the timing analysis,
spectral analysis was performed separately on three different data sets 
(a) the total time-averaged spectrum (``Phase T'' hereafter);
(b) a spectrum integrated over
the time intervals when the 5700~s binned 3--10~keV MECS
light curve has a count
rate ${\rm CR < CR_{break}}$ (``Phase A'' hereafter); (c) the
complement of Phase A (``Phase B'' hereafter).
Spectra from the three MECS units
were summed together after gain equalization and fit together
with the data from the other instruments.
Data were selected between 0.1--4.0~keV, 1.8--10~keV and
13--30~keV for the LECS, MECS and PDS, respectively.
Factors were included in the spectral fitting to account for 
known normalization uncertainties between the instruments
and the PDS to MECS factor was fixed to 0.75.
This value is a factor
$\simeq 0.82$ lower than reported
by Cusumano et al. (1998) for the Crab Nebula observation, to account
for the effects of
the RT screening algorithm. The following results are
not affected by a residual $\sim 10\%$ uncertainty on the exact value
of this parameter.
The spectral results are summarized in Table~\ref{tab2}.
\begin{table*}[hbt]
\caption{Best-fit spectral parameters}
\begin{center}
\begin{tabular}{lccccccc} \hline \hline
\multicolumn{8}{c}{Photoelectric absorbed power--law model} \\
Phase & ${\rm N_H}$ & ${\rm \Gamma}$ & $\chi^2/$~dof & & & & \\
& ($10^{20}$~cm$^{-2}$) & & & & & & \\ \hline
A & $3.83 \pm 0.11$ & $2.718 \pm 0.018$ & 839/450 & & & & \\
B & $3.84 \pm 0.08$ & $2.643 \pm 0.012$ & 1333/531 & & & &\\
T & $3.80 \pm^{0.06}_{0.05}$ & $2.660 \pm^{0.010}_{0.009}$ & 1833/561 & & & & \\ \hline \hline
\multicolumn{8}{c}{Photoelectric absorbed broken power--law model} \\
Phase & ${\rm N_H}$ & ${\rm \Gamma_1}$ & ${\rm \Gamma_2}$ & ${\rm E_{break}}$ & $\chi^2/$~dof & &  \\
& ($10^{20}$~cm$^{-2}$) & & & (keV) & & & \\ \hline
A & $1.9 \pm^{0.2}_{0.3}$ & $2.10 \pm^{0.10}_{0.09}$ & $2.82 \pm^{0.02}_{0.03}$ & $1.28 \pm^{0.13}_{0.11}$ & 474/448 & & \\
B & $2.03 \pm^{0.18}_{0.16}$ & $2.11 \pm^{0.06}_{0.07}$ & $2.734 \pm^{0.0015}_{0.013}$ & $1.39 \pm^{0.11}_{0.09}$ & 640/529 & & \\
T & $1.98 \pm^{0.13}_{0.12}$ & $2.11 \pm 0.05$ & $2.756 \pm^{0.0011}_{0.0012}$ & $1.35 \pm 0.07$ & 716/559 & & \\ \hline \hline
\multicolumn{8}{c}{Photoelectric absorbed broken power--law + 2 absorption edges model} \\
Phase & ${\rm N_H}$ & ${\rm \Gamma_1}$ & ${\rm \Gamma_2}$ & ${\rm E_{break}}$ & ${\rm E_{th}}$ & ${\rm \tau}$ & $\chi^2/$~dof \\
& ($10^{20}$~cm$^{-2}$) & & & (keV) & (keV) & &  \\ \hline
A & $1.7 \pm 0.2$ & $2.02 \pm 0.08$ & $2.96 \pm^{0.04}_{0.03}$ & $1.57 \pm^{0.11}_{0.13}$ & $1.25 \pm^{0.08}_{0.06}$ & $0.25 \pm^{0.09}_{0.10}$ & 425/445 \\
& & & & & 1.56$^{\dag}$ & $0.19 \pm^{0.11}_{0.12}$ & \\
B & $1.80 \pm^{0.17}_{0.15}$ & $2.00 \pm^{0.07}_{0.06}$ & $2.87 \pm0.03$ & $1.65 \pm^{0.13}_{0.10}$ & $1.26 \pm^{0.12}_{0.10}$ & $0.19 \pm 0.07$ & 531/525 \\
& & & & & $1.60 \pm^{0.18}_{0.05}$ & $0.22 \pm 0.08$ & \\
T & $1.77 \pm^{0.13}_{0.12}$ & $2.01 \pm 0.05$ & $2.88 \pm^{0.02}_{0.03}$ & $1.60 \pm^{0.04}_{0.12}$ & $1.23 \pm^{0.08}_{0.07}$ & $0.18 \pm^{0.06}_{0.05}$ & 561/554 \\
& & & & & $1.56 \pm 0.11$ & $0.22 \pm 0.06$ \\ \hline \hline
\multicolumn{8}{c}{Photoelectric absorbed variable curvature power--law model} \\
Phase & ${\rm N_H}$ & ${\rm \Gamma_E}$ & ${\rm \Gamma_H}$ & ${\rm E_0}$ & $\chi^2/$~dof & & \\
& ($10^{20}$~cm$^{-2}$) & & & (keV) & & & \\ \hline
\multicolumn{8}{c}{${\rm \beta = 1.0}$} \\
A & $1.8 \pm^{0.2}_{0.3}$ & $2.00 \pm^{0.10}_{0.12}$ & $2.73^{0.02}_{0.03}$ & $1.9 \pm 0.9$ & 436/447 & & \\
B & $2.04 \pm^{0.16}_{0.17}$ & $2.07 \pm 0.07$ & $2.66 \pm^{0.03}_{0.02}$ & $2.8 \pm 0.6$ & 553/529 & & \\
T & $1.93 \pm 0.13$ & $2.04 \pm 0.04$ & $2.664 \pm 0.014$ & $2.3 \pm 0.5$ & 601/558 & & \\ 
\multicolumn{8}{c}{${\rm \beta = 0.3}$} \\
A & $1.5 \pm^{0.2}_{0.3}$ & $1.3 \pm^{0.2}_{0.3}$ & $2.75^{0.04}_{0.03}$ & $2.9 \pm^{1.1}_{0.7}$ & 430/447 & & \\
& 1.64$^{\dag}$ & 1.45$^{\dag}$ & $2.77 \pm 0.03$ & $3.5 \pm^{0.5}_{0.4}$ & 431/449 & & \\
B & $1.74 \pm^{0.18}_{0.17}$ & $1.53 \pm^{0.14}_{0.15}$ & $2.70 \pm^{0.05}_{0.03}$ & $4.8 \pm^{2.8}_{1.2}$ & 529/529 & & \\
& 1.64$^{\dag}$ & 1.45$^{\dag}$ & $2.69 \pm 0.02$ & $4.2 \pm^{0.4}_{0.3}$ & 530/529 & & \\
T & $1.64 \pm^{0.14}_{0.13}$ & $1.45 \pm^{0.11}_{0.13}$ & $2.70 \pm 0.02$ & $3.7 \pm^{0.9}_{0.7}$ & 568/568 & & \\ \hline \hline
\end{tabular}
\end{center}
\noindent
$^{\dag}$fixed
\label{tab2}
\end{table*}

The X-ray spectra of BL Lac objects are generally well
described by a single power-law absorbed by an amount of matter
consistent with the Galactic value. Some BL Lac objects, however, seem to
require a model with a broken power-law, Mkn~421 being one of these
(Comastri et al. 1997; Takahashi et al. 1996).

\begin{figure}
\begin{center}
\epsfig{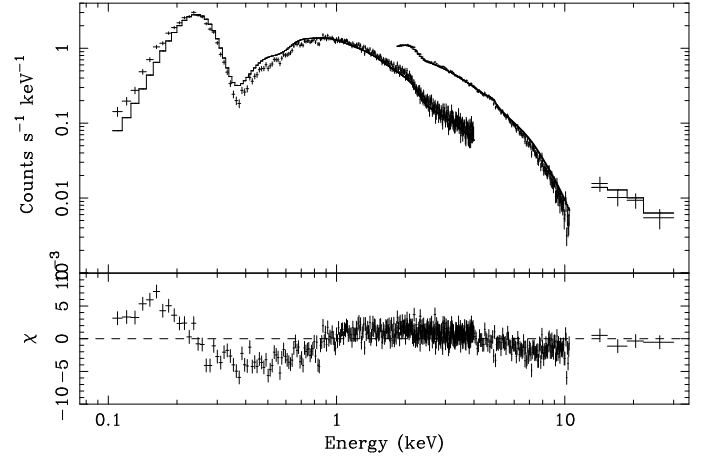}
\end{center}
\caption{Spectrum and best-fit model (upper panel) and residuals
in units of standard deviations (lower panel) when the time-averaged
spectrum of Mkn~421 is fit with a simple absorbed power-law
model. Each data point has a signal to noise ratio $>3$}
\label{fig4}
\end{figure}

A single power-law model provides an unacceptable fit for both the total
(reduced $\chi^2_{\nu}|_T = 3.26$) and the
intensity resolved spectra
($\chi^2_{\nu}|_A = 1.86$; $\chi^2_{\nu}|_B = 1.86$).
Fig.~\ref{fig4} demonstrates that this is not due to a localized
feature but to a incorrect modeling of the spectrum in the whole 0.1--30~keV
energy range. Although a broken power-law yields a dramatic improvement
of the quality of the fit, the fit is still unacceptable for
the T and B datasets
($\chi^2_{\nu}|_T = 1.28$; $\chi^2_{\nu}|_A = 1.06$;
$\chi^2_{\nu}|_B = 1.21$). A good fit can be obtained if a pair of
photoabsorption edges is added to the broken power-law,
with threshold energies
${\rm E^1_{th} \simeq 1.23}$~keV and ${\rm E^2_{th} \simeq 1.55}$~keV.
The addition of both edges is statistically required
in the time-averaged spectrum (${\rm \Delta \chi^2 = 126}$ and 29
for their subsequent inclusion in the spectral model). The best-fit
threshold energies are
broadly consistent with those expected from highly ionized Neon species
Ne{\sc ix} (${\rm E_{th} = 1.19}$~keV) and Ne{\sc x}
(${\rm E_{th} = 1.36}$~keV). No Oxygen edges are, however, detected, with the
90\% upper limit on the optical depth of an O{\sc vii} (O{\sc viii})
edge of 0.13 (0.12).
This makes such a model unlikely (see the discussion in Sect.~4),
despite an acceptable $\chi^2_{\nu}|_T = 1.04$.

If a narrow (i.e.: intrinsic width equal to 0)
Gaussian line is instead employed
to model the ``bump'' in the residuals around 1~keV, the improvement
in the quality of the fit is much less, albeit formally still highly
significant ($\Delta \chi^2 = 15$, corresponding to
chance occurrence likelihood of ${\rm \sim 1.5 \times 10^{-5}}$),
with best fit parameters: ${\rm N_H =
2.07\pm^{0.13}_{0.14} \times 10^{20}}$~cm$^{-2}$; ${\rm \Gamma_1
= 2.17\pm^{0.04}_{0.06}}$; ${\rm \Gamma_2 = 2.761\pm^{0.011}_{0.012}}$;
${\rm E_{break} = 1.51\pm^{0.12}_{0.16}}$~keV; ${\rm E_{line} =
0.97\pm^{0.05}_{0.04}}$~keV; ${\rm EW_{line} = 16\pm^6_7}$~eV.

An alternative explanation for the relatively poor fit of
the broken power-law model is that the intermediate X-ray spectrum
undergoes a gradual and smooth steepening with energy, which cannot
be described by an abrupt (and unphysical) switch in the spectral
photon index. A gradual steepening with energy on the other hand
agrees with the Synchrotron Self Compton (SSC) scenario, which is
nowadays widely accepted to explain the Spectral Energy Distribution (SED)
of BL Lac objects (Ghisellini, Maraschi \& Treves 1985; Ghisellini 1989).
We have therefore parameterized the gentle concave curvature
in the spectra with the function:
$$
{\rm F(E) = E^{ \{ -f(E) \Gamma_H + [1-f(E)] \Gamma_E \} } }
$$
where ${\rm f(E) = [1 - \exp(-E/E_0)]^{\beta}}$, ${\rm \Gamma_E}$
and ${\rm \Gamma_H}$ are the low and high energy asymptotic slopes
and ${\rm \beta}$ is a ``curvature radius'' in the energy space.
This model has only one degree of freedom more than the simple
broken power-law. This model has already successfully
fit the BeppoSAX spectrum of PKS~2155$-$304 (Giommi et al. 1998).
It was impossible to obtain a significant constraint on
${\rm \beta}$ from the fitting, the nominal best fit
value being $\simeq$0.30 for all datasets. In Table~\ref{tab2}
the results for ${\rm \beta = 0.3}$ and ${\rm \beta = 1}$ are shown.
The former case is strongly favored from the statistical point of
view and will be therefore discussed in the following. The $\chi^2_{\nu}$
is comparable with that from the broken power-law + edges model 
for all datasets
($\chi^2_{\nu} |_T = 1.02$; $\chi^2_{\nu} |_A = 0.96$;
$\chi^2_{\nu} |_B = 1.00$). The asymptotic spectral indices are
${\rm \Gamma_1 \simeq 1.45}$ and ${\rm \Gamma_2 \simeq 2.70}$,
with a folding energy ${\rm E_0 \simeq 3.7}$~keV. Interestingly,
the best-fit absorbing photoelectric column density (${\rm
N_H \simeq 1.64 \times 10^{20}}$~cm$^{-2}$) is 
well consistent with the Galactic contribution along the
line of sight to Mkn~421 (${\rm N_{H_{Gal}} = 1.6 \times 10^{20}}$~cm$^{-2}$,
Dickey \& Lockman 1990). The average 0.1--2~keV and 2--10~keV fluxes are
$1.76 \times 10^{-10}$~erg~cm$^{-2}$~s$^{-1}$ and 
$9.0 \times 10^{-11}$~erg~cm$^{-2}$~s$^{-1}$, corresponding to
rest frame luminosities of 7.4 and $3.9 \times 10^{44}$~erg~s$^{-1}$,
respectively.

\begin{figure}
\begin{center}
\epsfig{figure=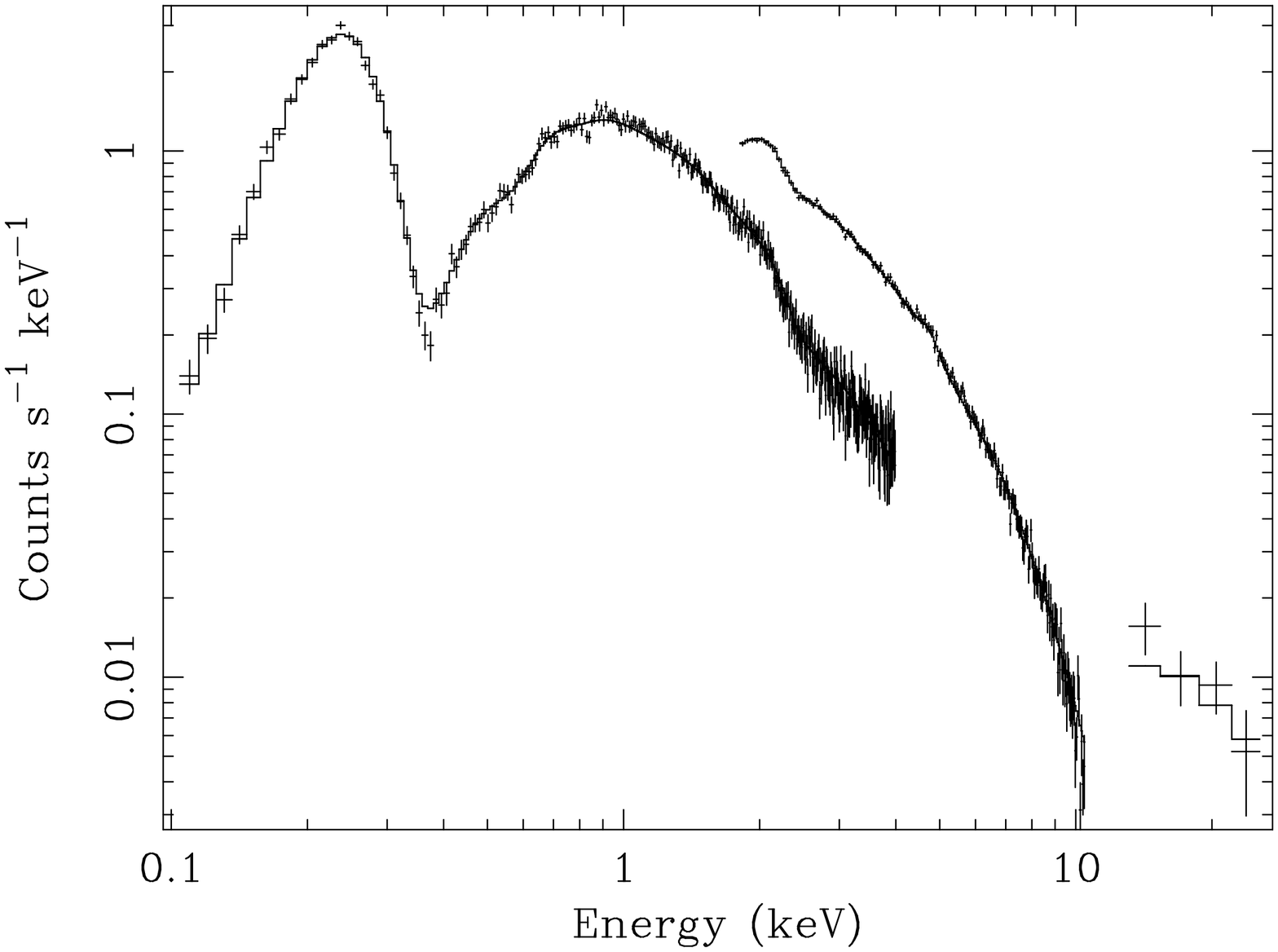,height=9.0cm,width=6.0cm,angle=-90}
\epsfig{figure=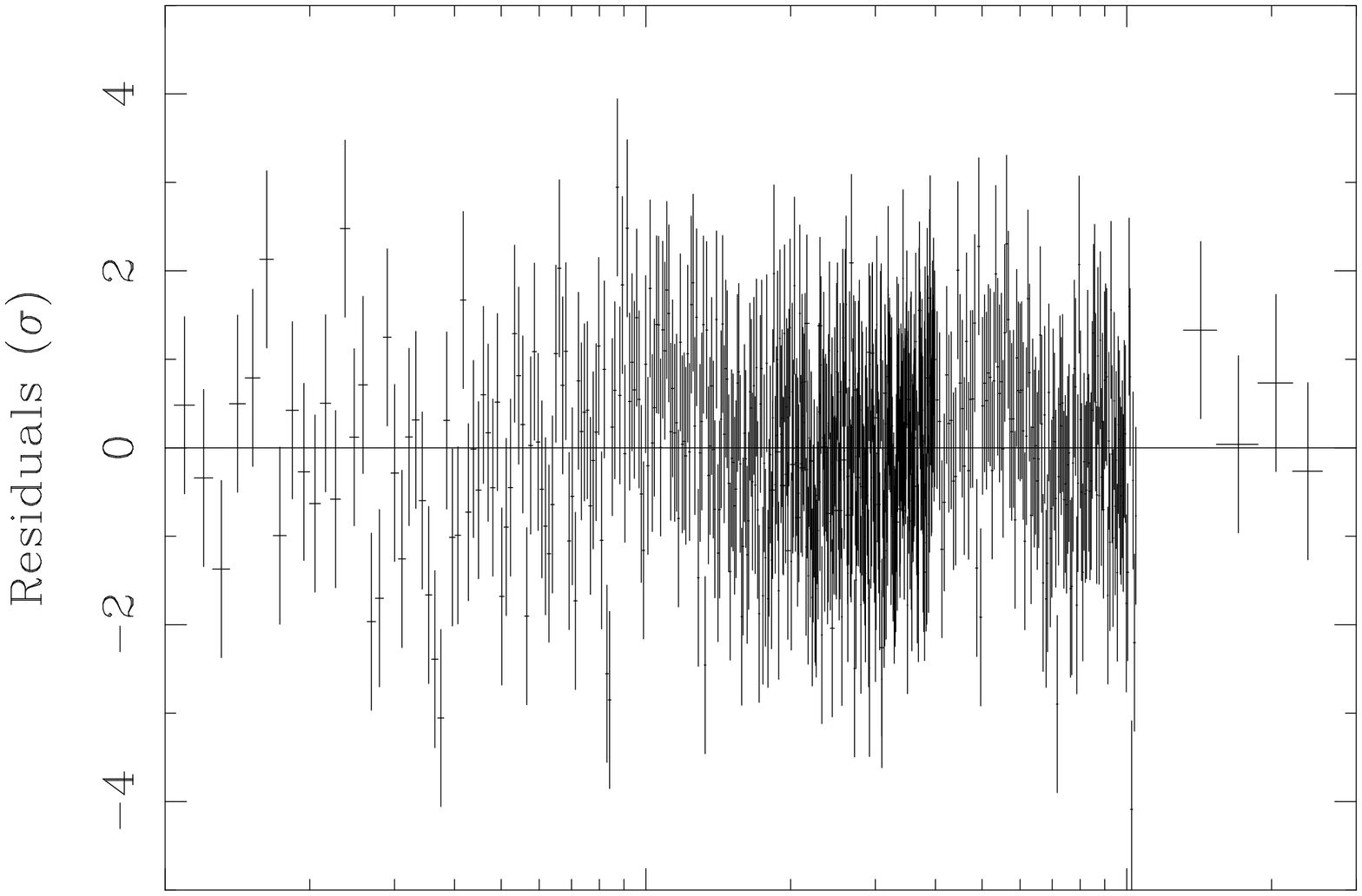,height=9.0cm,width=3.0cm,angle=-90}
\epsfig{figure=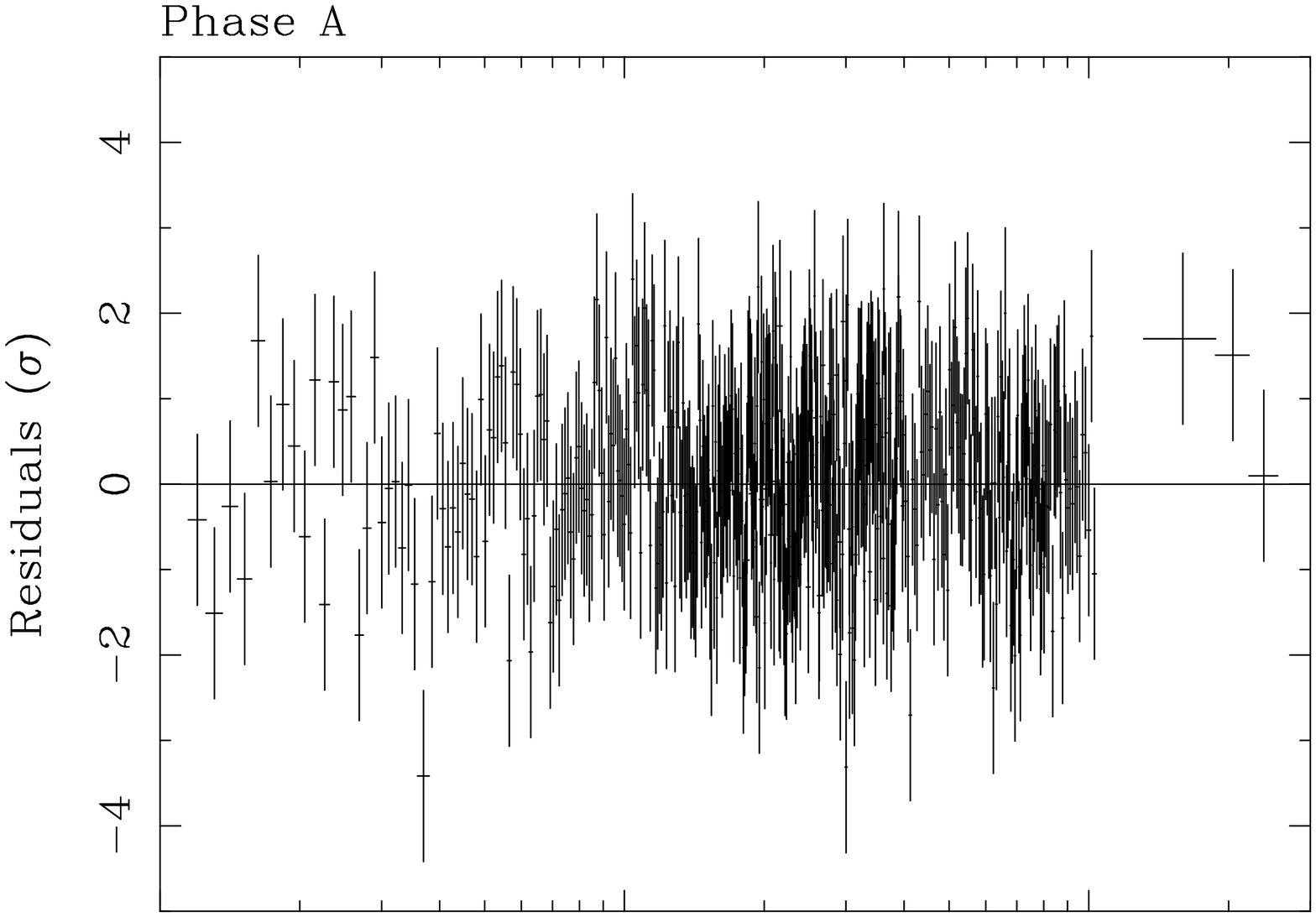,height=9.0cm,width=3.0cm,angle=-90}
\epsfig{figure=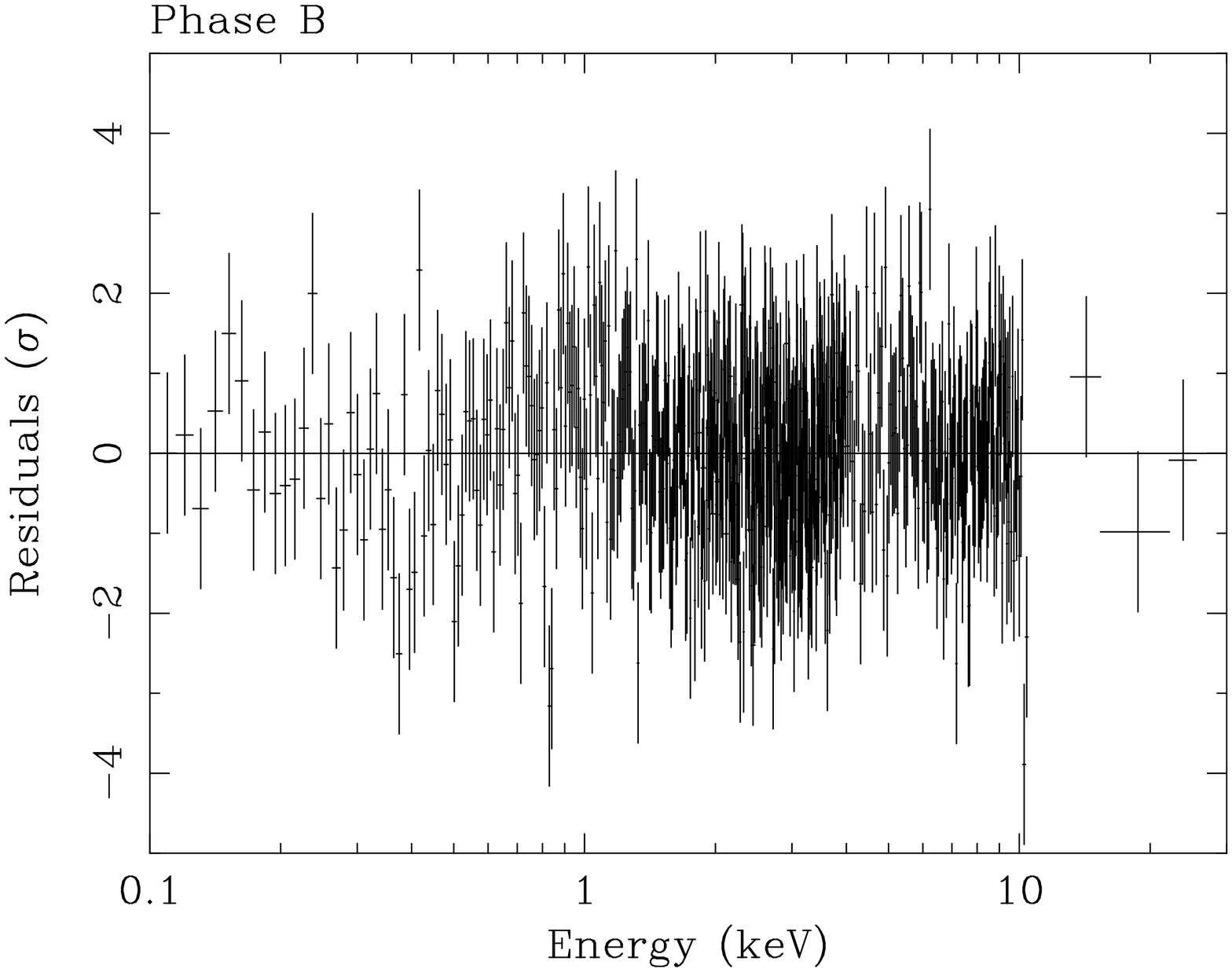,height=9.0cm,width=3.0cm,angle=-90}
\end{center}
\caption{Spectra and best-fit model when a photoelectric absorbed
power-law with a gradually changing spectral index is applied to
the time-averaged spectrum of all the instrument simultaneously
(upper panel).
The other panels show the residuals in units of standard
deviations when this model is applied to the time-averaged (upper middle),
Phase A (lower middle) and Phase B (lower) spectra}
\label{fig6}
\end{figure}

Using this physically reasonable description
of the continuum, both in the time averaged and intensity resolved
spectra, the next step is to understand which
model parameter(s) (and therefore which physical quantities) can
be considered responsible for the spectral changes observed in
Mkn~421. If all the parameters are left free, there is a suggestion
that ${\rm \Gamma_H}$ and ${\rm E_0}$ vary between Phase A and
B spectra, but they are still marginally consistent within the
statistical uncertainties. The fits on the intensity resolved spectra 
were therefore repeated, after fixing the other
two parameters (i.e.: ${\rm N_H}$ and ${\rm \Gamma_E}$) to their
best-fit values obtained from the time averaged spectrum fitting.
The asymptotic high energy slopes are different
at the 90\% confidence level, while the folding energies are still
consistent. However, whatever the detailed reason for the spectral
change is, it is what happens above $\simeq$4~keV that determines
the observed spectral variability. In Fig.~\ref{fig6}
the best-fit model and residuals are shown for all the datasets.

Narrow-band features such as Gaussian lines,
absorption edges or saturated absorption lines were added 
to the best-fit power-law with
variable curvature model. The largest improvement
in fit quality is obtained in the last case ($\Delta \chi^2
= 2.6$ for 2 degrees of freedom, which is significant at the 90\% level of
confidence only), with centroid energy ${\rm E_{notch} \simeq
0.55}$~keV. If the notch energy is held fixed at 0.654~keV
(L${\rm _{\alpha}}$ resonant absorption of O{\sc viii}),
following the discovery of such feature in the {\it
Einstein} spectra of several BL Lac objects (Canizares \& Kruper
1984; Madejski et al. 1991), only a 90\% upper limit of
10~eV on the EW can be set.

\section{Discussion}
\label{sec:disc}

Since the detection of this source at TeV energies (Punch et al.
1992), and the observation of correlation in the flaring activity at
X-ray and TeV energies, observations of Mkn~421 at X-rays have become
more important. The temporal coincidence of flaring in
X-ray and TeV supports emission models where both components are
produced by the same population of electrons with the X-rays at the
endpoint of the Synchrotron spectrum and the $\gamma$-rays at the
endpoint of an inverse Compton spectrum (Takahashi et al. 1996). This
is consistent with the overall picture drawn from the SED
of Mkn~421 which is characterized by two maxima, one in
the UV/soft X-ray band, and the second at GeV to TeV energies.

There is strong evidence that the X-ray
spectrum of High-energy peaked BL Lac objects (HBL)
is concave (Sambruna et al. 1994; Tashiro et al. 1995; Takahashi et al. 1996;
Sambruna et al. 1997). However,
even the ``standard'' broken power-law
model is unable to reproduce the spectral curvature of Mkn~421 observed here.
A system of two absorption edges, whose
threshold energies are broadly consistent with highly ionized
Neon species Ne{\sc ix} and Ne{\sc x},
is needed to get an acceptable fit quality. However, if the edges
represent the imprinting of a photoionized absorber along the line of
sight, it is difficult to explain the presence of highly ionized
Neon, while no Oxygen feature is detected. Assuming the
K-shell photoionization cross section formulae from Band et al. (1990),
the equivalent Neon column density inferred by the
measured optical depth is ${\rm N^{Ne}_H \sim 4 \times 10^{22}}$~cm$^{-2}$,
assuming that
the detected species represent the bulk of the elemental abundance.
In a photoionized plasma with spectral index $\Gamma \sim 2$, the
simultaneous presence of Ne {\sc ix} and Ne {\sc x} implies an
ionization parameter ${\rm \xi \sim 10^2}$, for which O{\sc viii}
should be the dominant Oxygen
ionic species (see e.g.
Kallman \& McCray 1982). Steeper spectra should only enhance the relative
Oxygen/Neon ratio. The observational upper limit on Oxygen abundance
is, however, more
than one order of magnitude lower than Neon (${\rm N^0_H < 2 \times
10^{21}}$~cm$^{-2}$). This scenario is viable only if deep
Oxygen edges were detected as well. However, ROSAT found no evidence
for these spectral features (Fink et al. 1991).
We conclude that such a description of the observed spectrum
is unphysical and can be discarded.

A good fit is obtained with
a model where the spectral steepening occurs gradually, hence no
additional spectral component is required. In this framework, the
hardening of the spectrum with increasing flux is basically due
to a change of the X-ray spectral properties above $\simeq$4~keV.

It is generally believed that the radiation
from BL Lac objects is relativistically beamed (Blanford \& Rees 1978) and
that the beam points directly towards the observer (Marscher 1980; K\"onigl
1981). 
The shape of the spectrum and the measurement of polarization in the
radio to optical bands provide the basis for the interpretation of the
emission in terms of SSC models. Here emission up
to X-ray wavelengths is caused by a relativistic population of
electrons via the synchrotron process, while the ${\rm \gamma}$-ray photons are
created by inverse Compton scattering of the same electron population
with the ambient photons
(Bregman et al. 1990;
Kawai et al. 1991). In the SSC model, the radiation output is driven by
the highest energy electrons. The observed curvature of the X-ray spectrum
can be explained by assuming that the jet is structured and
the electron distribution is located in a smaller (inner) region of
the jet with increasing energy. An increase in the synchrotron flux
means an extension of the region where the high energy electrons
are injected, probably in connection with increased activity of the
nuclear engine. Alternatively, the hardening of the spectrum
with increasing X-ray flux could be due to the injection of new
electrons in the jet, which are highly energetic and therefore
produce a flatter spectrum before they suffer significant radiative losses.

{\it Einstein} observations of a small sample of X-ray bright BL Lac
objects indicate that an absorption feature at 0.65~keV may be a
ubiquitous feature in their soft X-ray spectra (Majedski
et al. 1991).  This feature was first claimed to be detected by Canizares \&
Kruper (1984) in PKS~2155$-$304.
It was interpreted as Ly$_\alpha$ resonance absorption in
O{\sc viii} (${\rm E = 654}$~eV). However, ROSAT observations
(Fink et al. 1991) of
Mkn~421 found no evidence for such a spectral feature.
The BeppoSAX observation of PKS~2155$-$304 did not detect
such a feature either (Giommi et al. 1998),
casting further doubts on its
existence in BL Lac objects. In agreement with these findings,
our data show no evidence for any spectral features in the spectrum of
Mkn~421. In particular no absorption feature is observed around
0.6~keV, the 90\% upper limit on the equivalent width of
a Ly$_\alpha$ resonance absorption in O{\sc viii} being only 10~eV.

\section{Conclusions}

We summarize the main results of the presented BeppoSAX observations
of Mkn~421 as follows:

\begin{itemize}

\item[-] Mkn~421 was observed in a quiescent state, with an average 2--10~keV
flux of ${\rm 9.0 \times 10^{-11}}$~erg~cm$^{-2}$~s$^{-1}$. It exhibited
flux variation by a factor $\simeq$2 on timescales as short as a few $10^4$~s

\item[-] an energy threshold between different
variability patterns can be set at $\simeq$3--4~keV.
The 3--10~keV vs. 0.1--3~keV HR increases with increasing flux up to
$\simeq$$8 \times 10^{-11}$~erg~cm$^{-2}$~s$^{-1}$ and then ``saturates''
to a constant level

\item[-] the average 0.1--40~keV spectrum exhibits a gradual steepening
with energy. This effect is foreseen by the SSC scenario, where the X-rays
originate as synchrotron radiation from a relativistic
population of electrons in the jet.
If the proper (albeit admittedly phenomenological)
continuum model is adopted, no soft X-ray absorption feature is required,
with a 10~eV upper limit on the equivalent width of any ${\rm Ly_{\alpha}}$
resonance absorption line

\item[-] the observed spectral dynamics can be explained as an hardening of
the high-energy ({\it i.e.}: $\simgt$4~keV)
spectrum, above a relatively invariant soft X-ray emission. This can
be due to an increase of the injected electron flux or of
the extension of the injection region, possibly in connection with phases
of higher nuclear activity.

\end{itemize}

\begin{acknowledgements}

MG acknowledges an ESA research fellowship. BeppoSAX is a joint Italian-Dutch
program. The authors thank Dr. G. Matt for providing its {\sc Xspec}
gradually changing index power-law model.
Marco Beijersbergen made an invaluable contribution to the
determination of the LECS response by adapting his XMM ray-trace to the
BeppoSAX optics.

\end{acknowledgements}


\begin{thebibliography}{}

\bibitem{} Band I.M., Trzhaskovskaya M.B., Verner D.A., Yakovlev D.G., 1990, A\&A 237, 267

\bibitem{} Blandford R., Rees M.J., 1978, in ``Pittsburgh Conference on BL
Lacs Objects'', Wolfe A.M. ed. (Pittsburgh:Univ. Pittsburgh Press), 328

\bibitem{} Boella G., Perola G.C., Scarsi L., 1997a, A\&AS 122, 299

\bibitem{} Boella G., Chiappetti L., Conti G., et al., 1997b, A\&AS 122, 372

\bibitem{} Bregman J.N., Glassgold A.E., Huggings P.J., et al., 1990, ApJ 352, 574

\bibitem{} Canizares C.R., Kruper J., 1984, ApJL 278, 99

\bibitem{} Comastri A., Fossati G., Ghisellini G., Molendi S., 1997, ApJ 480, 534

\bibitem{} Cusumano G., Mineo T., Guainazzi M., et al., 1998, A\&A submitted

\bibitem{} Dickey J.M., Lockman F.J., 1990, ARA\&A 28, 215

\bibitem{} Fink H.H., Thomas H.C., Hasinger G., et al., 1991, A\&A 246, L6

\bibitem{} Frontera F., Costa E., Dal Fiume D., et al., 1997, A\&AS 122, 357

\bibitem{} George I.M., Warwick R.S., McHardy I.M., 1988, MNRAS, 235, 787

\bibitem{} Ghisellini G., 1989, MNRAS 236, 341

\bibitem{} Ghisellini G., Maraschi L., Treves A., 1985, A\&A 146, 204

\bibitem{} Giommi P., Barr P., Garilli B., Maccagni D., Pollock T., 1990, ApJ 356, 432

\bibitem{} Giommi P., Fiore F., Guainazzi M., et al., 1998, A\&A 333, 5

\bibitem{} Harmon B.A., Finger M.H., Rubin B., et al., 1992, Proceedings of the Compton Observatory Science Workshop, NASA CP3137, Shrader C.R., Gehrels N., Dennis B. eds., 69

\bibitem{} Kallman T., McCray R., 1982, ApJS 50, 263

\bibitem{} Kawai N., Matsuoka M., Bregman J.N., et al., 1991, ApJ 382, 508

\bibitem{} K\"onigl A., 1981, ApJ 243, 700

\bibitem{} Kubo H., Takahashi T., Madejski G., et al., 1998, ApJ 504, 693

\bibitem{} Lampton M., Margon B., Bowyer S., 1976, ApJ 208, 177
 
\bibitem{} Lin Y.C., Bertsch D.L., Dingus B.L., et al., 1996, A\&AS 120, 499

\bibitem{} Macomb D.J., Akerlof C.W., Aller H.D., et al., 1995, ApJ 449, 99

\bibitem{} Madejski G.M., Mushotzky R.F., Weaver K.A., Arnaud K.A., 1991,
ApJ 370, 198

\bibitem{} Manzo G., Giarrusso S., Santangelo A., et al., 1997, A\&AS 122, 341

\bibitem{} Marscher A.P., 1980, ApJ 235, 386

\bibitem{} Parmar A., Martin D.D.E., Bavdaz N., et al., 1997, A\&AS 122, 309

\bibitem{} Punch M., Akerlof C.W., Cawley M.F., et al., 1992, Nat 358, 477

\bibitem{} Ricketts M.J., Cooke B.A., Pounds K.A., 1976, Nat 259, 546

\bibitem{} Sambruna R.M., Barr P., Giommi P., et al., 1994, ApJS 95, 317

\bibitem{} Sambruna R.M., George I.M., Madejski G., et al., 1997, ApJ 482, 774

\bibitem{} Tashiro M., Makishima K., Ohashi T., et al., 1995, PASJ 47, 131

\bibitem{} Takahashi T., Tashiro M., Madejski G., et al., 1996, ApJ 470, 89

\bibitem{} Ubertini P, Bazzano A., La Padula C., Polcaro V.F., Manchanda R.K., 1984, ApJ 284, 54

\end{thebibliography}
\end{document}